# Monitoring of food spoilage by high resolution THz analysis

Francis Hindle[1], Lotta Kuuliala[2], Meriem Mouelhi[1,3], Arnaud Cuisset[1], Cédric Bray[1], Mathias Vanwolleghem[4], Frank Devlieghere[2], Gaël Mouret[1], and Robin Bocquet[1]

[1]Laboratoire de Physico-Chimie de l'Atmosphère (LPCA EA 4493), Université du Littoral Côte d'Opale, 189A Av. Maurice Schumann, 59140 Dunkerque, France

[2]Research Unit Food Microbiology and Food Preservation (FMFP), Department of Food Technology, Safety and Health, Part of Food2Know, Faculty of Bioscience Engineering, Ghent University, Coupure links 653, Ghent, Belgium

[3]Laboratoire Dynamique Moléculaire et Matériaux Photoniques, Ecole Nationale Supérieure d'Ingénieurs de Tunis, Université de Tunis – 5 av. Taha Hussein, 1008 Tunis, Tunisia

[4]Institute of Electronics, Microelectronics and Nanotechnology, CNRS, Centrale Lille, ISEN, Université de Valenciennes, UMR 8520, University of Lille 1, Villeneuve d'Ascq, France

**Abstract**

High resolution rotational Terahertz (THz) spectroscopy has been widely applied to the studies of numerous polar gas phase molecules, in particular volatile organic compounds (VOCs). During the storage of foodstuffs packed under a protective atmosphere, microbial activity will lead to the generation of a complex mixture of trace gases that could be used as food spoilage indicators. Here we have demonstrated that the THz instrumentation presently available provides sufficient sensitivity and selectivity to monitor the generation of hydrogen sulfide ($H_2S$) in the headspace of packed Atlantic salmon (*Salmo salar*) fillet portions. A comprehensive comparison was made by selective-ion flow-tube mass spectrometry (SIFT-MS) in order to validate the THz measurements and protocol. The detectivity of a range of alternative compounds for this application is also provided, based on the experimental detection limit observed and molecular spectroscopic properties. Molecules like ethanol, methyl mercaptan and ammonia are suitable indicators with the presently available sensitivity levels, while dimethyl sulfide, acetone and butanone may be considered with a sensitivity improvement of 2 orders of magnitude.

## 1. Introduction

Food spoilage is one of the most fundamental challenges of today's society. In the industrialized countries, over 40 % of food produced annually has been estimated to be wasted at retail and consumer levels [1]. Especially in the case of highly perishable food products such as seafood, quality concerns can lead to an increase in waste levels due to precautionary measures being taken. Hence, the ability to confidently determine the spoilage status is of upmost importance for optimizing the quality and economic viability of food products.

Spoilage of fresh food products is typically due to microbial activity, a consequence of which is the production of numerous volatile compounds. Trace quantities of these molecules accumulate in the package headspace and commonly leads to the generation of unpleasant off-odors that eventually cause consumer rejection. Monitoring of spoilage-related volatile compounds thus provides several possibilities for the development of intelligent packaging technologies [2]–[5]. For example, production of sulfuric compounds has often been observed during spoilage processes of seafood [6]–[10] and vegetables [11]–[13]. Hydrogen sulfide ($H_2S$) is produced by many common spoilage bacteria [14], [15] and has previously been applied for monitoring the quality of muscle foods [16], [17]. Many small polar molecules display a strong rotational spectral fingerprint in the THz domain. The analysis of a material by THz spectroscopy is particularly attractive as it has great potential for the discrimination between different chemical species, isomers [18], conformers [19], and isotopologues [20]. Such a degree of selectivity is obtained for the analysis of gas phase samples and especially at low pressure when the line widths are not broadened by pressure effects. These properties have already been demonstrated for the analysis of cigarette smoke [21] and other complex mixtures of gases [22]. In this study we demonstrate the potential of THz for the monitoring of food spoilage. THz spectroscopy was used for measuring different $H_2S$ levels in the headspace of Atlantic salmon (*Salmo salar*) packaged under 100 % $N_2$. The obtained results were validated by comparison with selective-ion flow-tube mass spectrometry (SIFT-MS).

## 2. THz spectrometer

High frequency monochromatic radiation sources operating at frequencies up to 1 THz can be realized by cascading a series of non-linear elements each optimized for a given band and frequency multiplication rank. The first multiplier is driven by a standard microwave synthesizer generally operating in the range of 8 GHz to 18 GHz. In our case a combination of seven different multipliers are used to cover the frequency range from 70 GHz to 900 GHz. The power delivered by the source is strongly dependent on frequency, caused by the number of cascaded elements and the low conversion efficiency of each element particularly at high frequency.

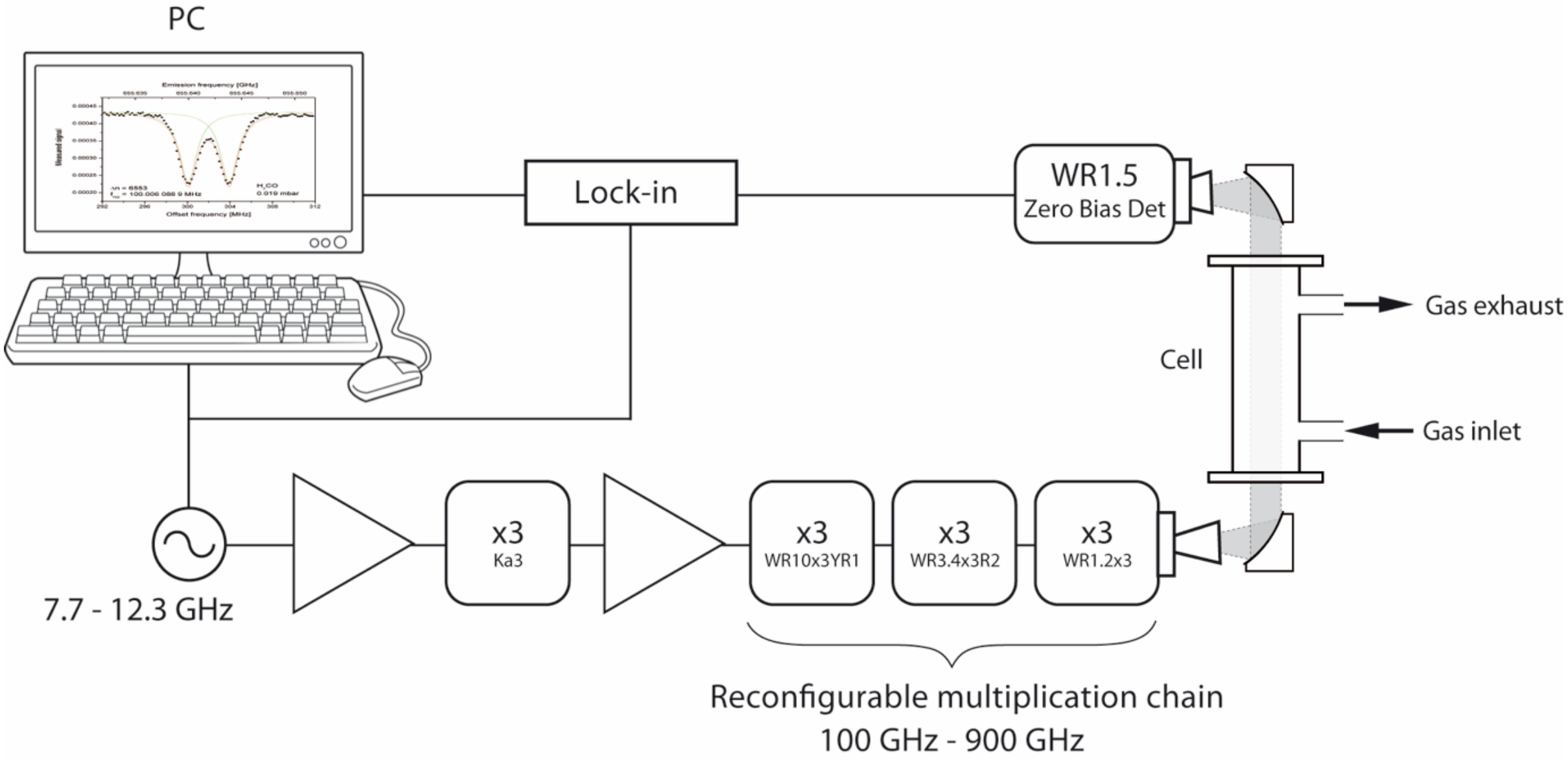

**Figure 1. Frequency multiplication spectrometer shown with an overall multiplication rank of × 81 able to cover 630 GHz to 900 GHz.**

A high-resolution THz spectrometer can be readily constructed using such an amplified frequency multiplication chain [23]. The waveguided radiation is launched into free space using a suitable horn, before being propagated through a measurement cell and focused onto a detector by means of parabolic mirrors, figure 1. A cryogenically cooled bolometer can be used for detection, however for operation at room temperature a GaAs Schottky diode can be employed with excellent results. Such an instrument has a number of appealing characteristics, excellent frequency traceability and narrow linewidth emission. Entirely composed of commercially available components the instrument construction is straightforward and readily automated for an autonomous data acquisition. The accuracy of the emission frequency is critical for line position measurements, here the base synthesizer is directly referenced to a known frequency signal. The emission linewidth depends on the phase noise of the reference oscillator, reasonable precautions allow position measurements with an uncertainty of 10 kHz to be obtained. Both Amplitude Modulation (AM) and Frequency Modulation (FM) schemes can be employed, independently or even simultaneously with the desired parameters being extracted from the detector signal by a lock-in amplifier. The computer control of both the microwave synthesizer and the lock-in allows complete automation of the data acquisition process. The relatively straightforward operation of this instrument allowed it to be applied the high-resolution rotational spectroscopy of a wide variety of polar molecular systems and the detection trace gases in industrial emissions [24]. Indeed, the FM scheme was used to increase the signal to noise ratio of the molecular signals generated by weakly absorbing lines, improving the trace detection limit. In FM mode the measured signal is sensitive to the gradient of the AM spectrum, and the strength of the response not only depends on molecular signal in question but also parameters like detector sensitivity, modulation depth, etc. The measured FM spectrum alone does not contain sufficient information to permit the direct quantification of the target species. Nevertheless, comparison of the response to a calibrated mixture allows sensitive quantification to be undertaken. In principle the spectrometer sensitivity is only limited by the source power and the detector noise. The instrument may also be limited by the stability of the baseline which contains strong Fabry-Pérot interference. The measurement of a variety of well-known gases and in particular molecular transitions approaching the sensitivity limit allow the smallest detectable change in the absorption coefficient to be estimated.

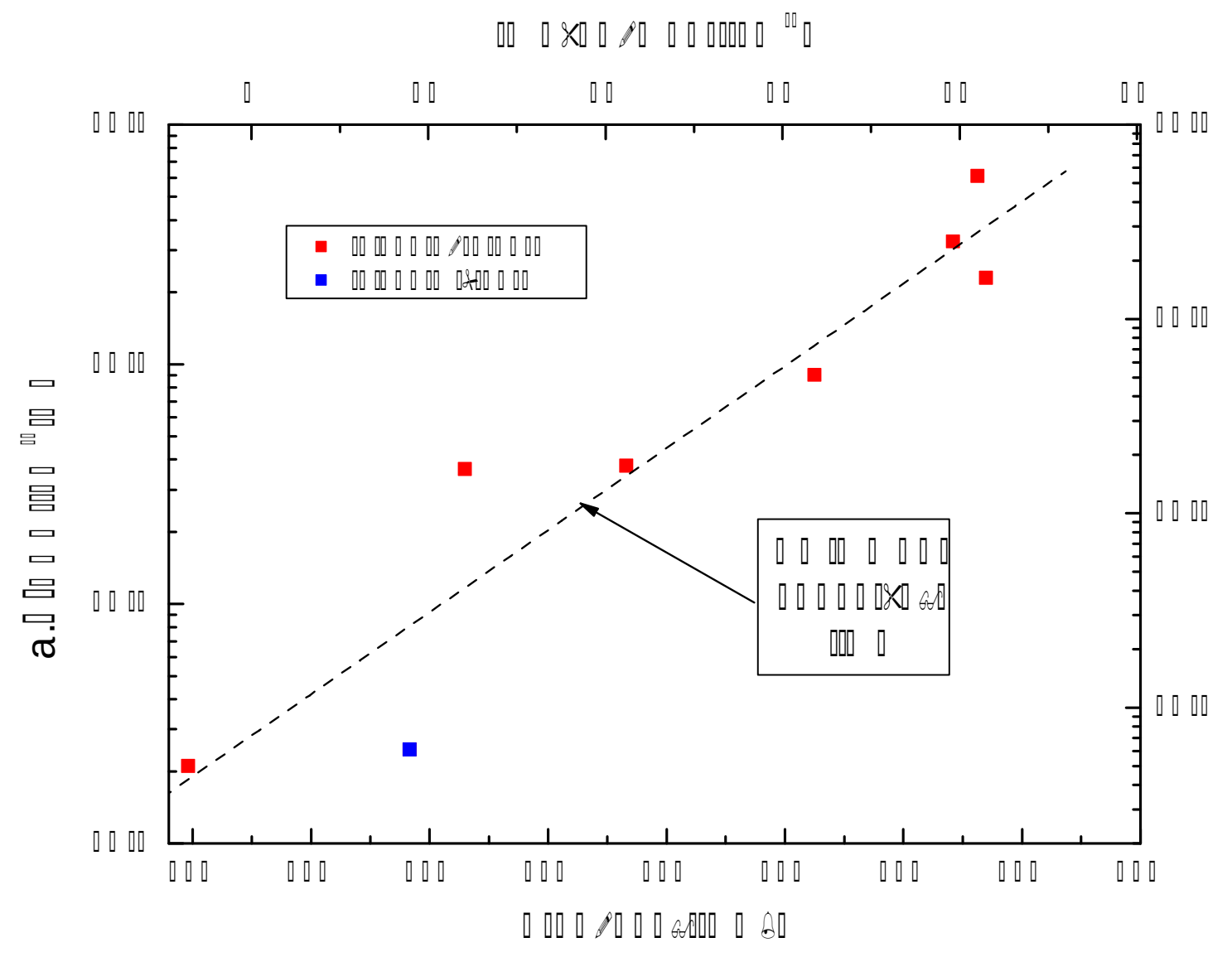

**Figure 2. Estimation of the minimum detectable absorption coefficient $\alpha_{min}$ as a function of frequency for an interaction length of 1m. All values are for a measurement bandwidth of 1 Hz.**

**Red squares for pure gases, and a blue square for a calibrated mixture of $SO_2$ in nitrogen at 100 ppm. Room temperature operation with zero bias Schottky diode detector.**

The minimum detectable absorption coefficient $\alpha_{min}$ has been experimentally evaluated over the frequency range from 100 GHz to 800 GHz, figure 2. Several molecular lines of pure gases have been measured, $C_2H_5OH$, $CH_3CN$ and $NH_3$, providing line strengths from $5 \times 10^{-24}$ to $3 \times 10^{-22}$ ($cm^{-1}$/(molecule.$cm^{-2}$)). In addition, a calibrated mixture containing a trace $SO_2$ diluted in nitrogen was also used, in this case with a mixing ratio ($\chi$) of 100 parts per million (ppm). The maximum absorption coefficient $\alpha_0$ of the molecular lines encountered at the line centers was determined by equations 1 and 2 with the line strength (S), and broadening coefficients ($\gamma_{self}$ and $\gamma_{air}$), tabulated in spectroscopic databases [25], [26].

$$\alpha_{0\,pure} = \frac{S}{k_B T_0 \pi \gamma_{self}} \tag{1}$$

$$\alpha_{0\,trace} = \frac{S\chi}{k_B T_0 \pi \gamma_{air}} \tag{2}$$

where $k_B$ is the Boltzmann constant and $T_0$ the temperature. The values of molecular absorption coefficient $\alpha_0$ are normalized for a path length of 1 m and SNR = 1, leading to the quantity of $\alpha L/SNR$. Although this quantity is dimensionless here it is established with the usual units yielding a scale in [$cm^{-1}$.m] which may be directly interpreted. The SNR is that of the molecular signal, defined as the ratio of the line absorption profile amplitude, in AM or FM, to the noise amplitude of the background. A molecular transition with a SNR = 1 is the weakest line that can be distinguished from the baseline noise, and is considered to represent the sensitivity limit of the instrument. The values determined here are for a measurement bandwidth $\Delta f$ = 1 Hz, they are therefore directly equivalent to the Noise Equivalent Absorption (NEA) in units of $cm^{-1}$ .$Hz^{-1/2}$ for a path length of 1 m .The minimum detectable absorption coefficient displays a strong power law dependency with the frequency, its value increasing by one decade for each step of approximately 300 GHz. A trend line is included in figure 2 to assist the interpretation but does not suggest a rigorous solution. The frequency dependency is a direct result of the available power and detector NEP which deteriorate as the frequency is increased. In the case of $H_2S$, selecting only the most intense transitions yield an expected detection limit in the order of 500 ppb for an interaction length of 1 m.

## 3. Preparation of salmon samples

Norwegian farmed Atlantic salmon was gutted, filleted and skinned at a commercial seafood processing company and delivered to the Research Unit Food Microbiology and Food Preservation (FMFP, Ghent, Belgium) four days after harvest in polystyrene boxes, wrapped in plastic and covered with ice. Five fillets A-E (each ca. 1.2-1.5 kg) were manually mixed for one minute to ensure uniform surface contamination, portioned (203.0 ± 2.0 g) and packed under 100 % $N_2$ with a 2:1 gas-product ratio, using tray-sealer MECA 900 (DecaTechnic, Herentals, Belgium), trays (PP/EVOH/PP, $O_2$ transmission rate 0.001 $cm^3$/(tray • 24 h) at 23 °C and 50 % RH) and top film (PA/EVOH/PA/PP, $O_2$ transmission rate 6.57 $cm^3$/($m^2$ • 24 h • atm) at 23 °C, 50 % RH and 1 atm). Six individual packages were prepared from each fillet and stored at 4.0 ± 0.7 °C for up to 13 days. Initial SIFT-MS analyses (section 4) were performed on a regular basis (day 1, 3, 5, 7, 9, 11 and 13, where day 0 is the day of packaging) in order to estimate the $H_2S$ concentrations for the THz measurements (section 5). On each day, 3-6 randomly selected packages were analyzed; after analyses the fillet portions were individually re-packed in high barrier bags ($O_2$ transmission rate < 2.7 $cm^3$ / ($m^2$ • 24 h • 1 bar) at 23 °C and 0 % RH) under vacuum and stored at -32 °C for up to 22 days. In this way, multiple samples representing different storage times and $H_2S$ levels were available for measurement by THz spectroscopy.

Comparative THz spectroscopy and SIFT-MS analyses were carried out for selected samples presenting $H_2S$ concentrations between 400 to 4000 ppb. The selected samples were thawed overnight at 2 °C, re-packaged in trays under 100 % $N_2$ and stored at 4 °C for > 1 h before the analysis. Each sample was measured by THz spectroscopy and SIFT-MS simultaneously. A 2 ml volume of the headspace gas was extracted from the sample and its THz spectrum recorded. During the measurement of the THz spectrum the sample was analyzed by SIFT-MS located in the same building. The entire process for a single sample was completed in approximately 15 minutes and a total of ten samples were measured (1x5days, 1x7days, 3x9days, 4x11days, 1x13days).

## 4. Selective-ion flow-tube mass spectrometry (SIFT-MS)

SIFT-MS is a mass spectrometry technique that can be used for real-time quantification of trace gases and has been previously validated for seafood spoilage analysis [27]. The method is based on soft chemical ionization of trace gases by precursor ions (typically $H_3O^+$, $NO^+$, $O_2^+$; these ions react with a wide variety of trace gases but not with air components). Precursor ions are produced in a discharge ion source and selected to a current of ions with known mass-to-charge (m/z) ratios by a quadrupole mass filter. The selected precursor ions are injected in a flowing carrier gas with a controlled flow rate and react with the trace gases of the sample gas. The resulting product ions and the remaining precursor ions are separated on the basis of their m/z ratio and counted. Using the count rates of product and precursor ions and the corresponding reaction rate coefficient (k), the concentration of the trace gases in the sample gas can be determined [28], [29].

In this study, the spectrometer (Voice 200, Syft Technologies, Christchurch, New Zealand) was used in Multiple Ion Monitoring (MIM) mode for quantifying 25 species from the package headspace (Table A) in accordance with the principles presented by *Sader et al.* [30]. Prior analyses, the flow rate was measured with a soap film flowmeter (Gilibrator-2, Sensidyne, St. Petersburg, FL, USA; glass bubble flowmeter, Agilent Technologies, Santa Clara, CA, USA) and all measured concentrations were corrected accordingly. Sampling was carried out through a septum placed on the top film; the package headspace was simultaneously connected to a 100 % $N_2$ bag in order to avoid package collapse. The package headspace was sampled for ca. 300 s and concentrations were averaged over the obtained datapoints (ca. 40). Relative standard deviations ($SD_{\%}$) were determined for each sample, ($SD_{\%} = SD_m/x_m * 100$ %, where $x_m$ is the average and $SD_m$ the standard deviation of a given molecule). Product ions used in quantification (Table A) were selected by optimizing following consecutive criteria: 1) $SD_{\%}$ < 25 %, 2) high branching ratio, 3) minimum amount of m/z conflicts, 4) lowest concentration.

**Table A. The SIFT-MS method: Trace gas, precursor and product ions, mass-to-charge ratios (m/z), reaction rates (k) and branching ratios (b). Product ions marked with (*) were selected for quantification.**

| Trace gas | Precursor | m/z | b (%) | k | Product ion |
|---|---|---|---|---|---|
| **Acids** | | | | | |
| Acetic acid | $NO^+$ | 90 | 100 | 9.0 E -10 | $NO^+.CH_3COOH^*$ |
| | $NO^+$ | 108 | | 9.0 E -10 | $NO^+.CH_3COOH.H_2O$ |
| 3-methylbutanoic acid | NO+ | 132 | 70 | 2.5 E -09 | $C_5H_{10}O_2.NO^{+*}$ |
| **Alcohols** | | | | | |
| 2,3-butanediol | $NO^+$ | 89 | 100 | 2.3 E -09 | $C_4H_9O_2^{+*}$ |
| | $NO^+$ | 107 | | 2.3 E -09 | $C_4H_9O_2^+.H_2O$ |
| Ethanol | $NO^+$ | 45 | 100 | 1.2 E -09 | $C_2H_5O^{+*}$ |
| | $NO^+$ | 63 | | 1.2 E -09 | $C_2H_5O^+.H_2O$ |
| | $NO^+$ | 81 | | 1.2 E -09 | $C_2H_5O^+.2(H_2O)$ |
| 3-methyl-1-butanol | $H_3O^+$ | 71 | 100 | 2.8 E -09 | $C_5H_{11}^{+*}$ |
| | $NO^+$ | 87 | 85 | 2.3 E -09 | $C_5H_{11}O^+$ |
| Isobutyl alcohol | $NO^+$ | 73 | 95 | 2.4 E -09 | $C_4H_9O^{+*}$ |
| | $O_2^+$ | 33 | 50 | 2.5 E -09 | $CH_5O^+$ |
| **Aldehydes** | | | | | |
| 3-methylbutanal | $NO^+$ | 85 | 100 | 3.0 E -09 | $C_5H_9O^{+*}$ |
| **Aromatic hydrocarbons** | | | | | |
| Ethyl benzene | $NO^+$ | 106 | 100 | 2.0 E -09 | $C_8H_{10}^{+*}$ |
| Propyl benzene | $NO^+$ | 120 | 100 | 2.0 E -09 | $C_9H_{12}^{+*}$ |
| Styrene | $NO^+$ | 104 | 100 | 1.7 E -09 | $C_8H_8^{+*}$ |
| **Ketones** | | | | | |
| Acetone | $NO^+$ | 88 | 100 | 1.2 E -09 | $NO^+.C_3H_6O^*$ |
| Acetoin | $NO^+$ | 118 | 100 | 2.5 E -09 | $C_4H_8O_2.NO^{+*}$ |
| 2,3-butanedione | $NO^+$ | 86 | 65 | 1.3 E -09 | $C_4H_6O_2^{+*}$ |
| Butanone | $NO^+$ | 102 | 100 | 2.8 E -09 | $NO^+.C_4H_8O^*$ |
| **Sulfur compounds** | | | | | |
| Carbon disulfide | $O_2^+$ | 76 | 100 | 7.0 E -10 | $CS_2^{+*}$ |
| Dimethyl sulfide | $H_3O^+$ | 63 | 100 | 2.5 E -09 | $(CH_3)_2S.H^{+*}$ |
| | $NO^+$ | 62 | 100 | 2.2 E -09 | $(CH_3)_2S^+$ |
| | $O_2^+$ | 47 | 25 | 2.2 E -09 | $CH_3S^+$ |
| | $O_2^+$ | 62 | 60 | 2.2 E -09 | $(CH_3)_2S^+$ |
| Dimethyl disulfide | $H_3O^+$ | 95 | 100 | 2.6 E -09 | $(CH_3)_2S_2.H^+$ |
| | $NO^+$ | 94 | 100 | 2.4 E -09 | $(CH_3)_2S_2^{+*}$ |
| Dimethyl trisulfide | $H_3O^+$ | 127 | 100 | 2.8 E -09 | $C_2H_6S_3H^+$ |
| | $H_3O^+$ | 145 | | 2.8 E -09 | $C_2H_6S_3H^+.H_2O$ |
| | $NO^+$ | 126 | 100 | 1.9 E -09 | $C_2H_6S_3^{+*}$ |
| Hydrogen sulfide | $H_3O^+$ | 35 | 100 | 1.6 E -09 | $H_3S^{+*}$ |
| | $H_3O^+$ | 53 | | 1.6 E -09 | $H_3S^+.H_2O$ |
| Methyl mercaptan | $H_3O^+$ | 49 | 100 | 1.8 E -09 | $CH_4S.H^{+*}$ |
| | $H_3O^+$ | 67 | | 1.8 E -09 | $CH_4S.H^+.H_2O$ |
| **Esters** | | | | | |
| Ethyl acetate | $H_3O^+$ | 89 | 100 | 2.9 E -09 | $CH_3COOC_2H_5.H^{+*}$ |
| | $H_3O^+$ | 107 | | 2.9 E -09 | $CH_3COOC_2H_5.H^+.H_2O$ |
| | $O_2^+$ | 31 | 20 | 2.4 E -09 | $CH_3O^+$ |
| | $O_2^+$ | 61 | 40 | 2.4 E -09 | $C_2H_5O_2^+$ |
| **Amines** | | | | | |
| Ammonia | $H_3O^+$ | 18 | 100 | 2.6 E -09 | $NH_4^+$ |
| | $H_3O^+$ | 36 | | 2.6 E -09 | $NH_4^+.H_2O$ |
| | $O_2^+$ | 17 | 100 | 2.6 E -09 | $NH_3^{+*}$ |
| Dimethyl amine | $H_3O^+$ | 46 | 100 | 2.1 E -09 | $(CH_3)_2NH.H^{+*}$ |
| Piperidine | $H_3O^+$ | 86 | 90 | 3.4 E -09 | $C_5H_{12}N^{+*}$ |
| Trimethyl amine | $NO^+$ | 59 | 100 | 1.6 E -09 | $(CH_3)_3N^{+*}$ |

## 5. Hydrogen sulfide quantification by THz spectroscopy

Hydrogen sulfide presents many absorption lines in the THz domain, approximately 200 lines are tabulated in the spectroscopic databases with line strengths from in the order of $10^{-23}$ to $10^{-20}$ ($cm^{-1}$/(molecule.$cm^{-2}$)). The selection of a suitable absorption line for trace detection is compromise between the line strength, available power and presence of other molecular lines such as $H_2O$. In our case the line at 611.441 GHz with a line strength of 4.18 x $10^{-21}$ ($cm^{-1}$/(molecule.$cm^{-2}$)) was selected in preference to the strongest line at 736 GHz. Although the line at 736 GHz has twice the line strength the power available from our source at this frequency approximately 4 times less. The amplified multiplier chain was configured with a rank of $\times$ 54, covering the frequencies from 440 GHz to 660 GHz and providing between 10 to 36 $\mu$W of power across this band.

In order to achieve the maximum sensitivity, a FM scheme was used with the detection being performed at the second harmonic of the modulation frequency. The modulation depth was fixed at 1188 kHz close to the Doppler linewidth at room temperature for this transition. Using this configuration, the recorded signal is sensitive to the gradient of the absorption profile, hence a narrow will provide the best contrast. A calibration gas containing $H_2S$ diluted in nitrogen with a mixing ratio of 50 ppm was measured to establish the response of the instrument, figure 3. At low pressure the line is Doppler limited hence the linewidth is constant and the absorption will increase linearly with the pressure, as shown in the second pane of figure 3 for pressures from 0.0 to 0.1 mbar. Above this limit the linewidth begins to increase due to the air broadening of the line. The maximum signal is observed when the collisional broadening width is equal to the Doppler linewidth, 0.25 mbar in our case. Above this pressure the linewidth increases with no change in the maximum absorption coefficient, the FM signal therefore decreases.

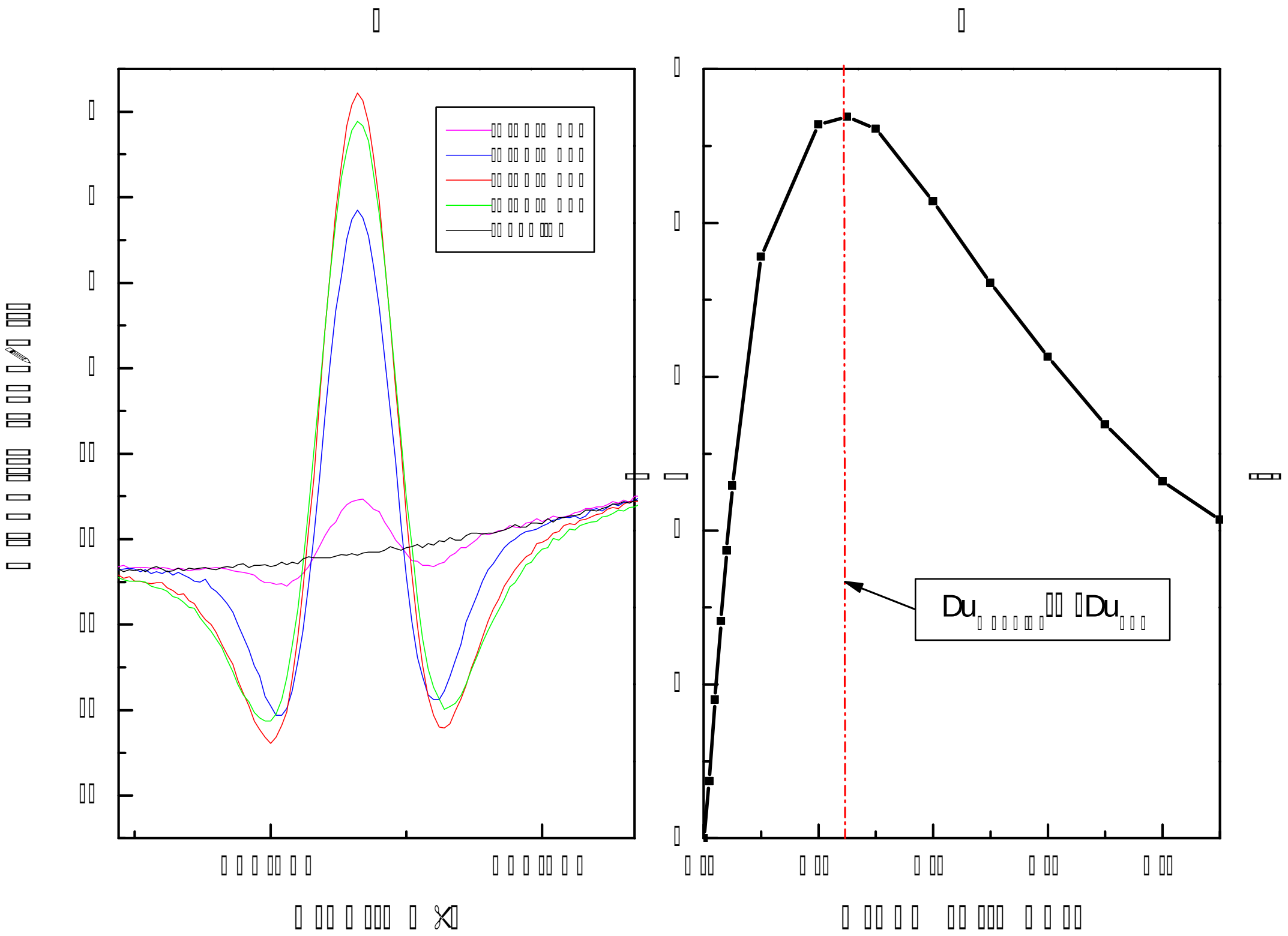

**Figure 3. Left pane, spectra of a calibrated 50 ppm trace $H_2S$ in $N_2$, recorded with a modulation depth of 1188 kHz, an interaction length of 1.128 m, and at a temperature of 293 K. Right pane, signal amplitude as a function of the working pressure, maximum value when the collisional broadening matches the Doppler linewidth.**

The sample cell was composed of a stainless-steel tube of length 1.128m with an internal diameter of 50 mm. The ends were closed by Teflon windows allowing the THz radiation to enter and exit the measurement cell. A turbo vacuum pump, pressure gauge and septum were also connected to the cell. A gas sample was extracted from the headspace of the salmon fillet portions packed in trays via a septum placed on the top film. A volume of 2 ml of the headspace gas was taken at atmospheric pressure using a gas tight syringe, this was injected into the evacuated sample cell (pressure < $10^{-2}$ mbar) typically yielding a pressure of approximately 0.40 mbar in the measurement cell. The pressure was reduced to obtain the optimal of 0.25 mbar and the THz spectra recorded.

A model was required to enable the correct extraction of the desired information from the recorded spectra. As the optimal signal is obtained when the collisional broadening equals the Doppler linewidth a model was sought to account for both of these contributions. A Fast Fourier Transform method was selected and is capable of calculating the response of a spectral line in this Voigt regime when measured in FM mode [31]. The measured molecular signals are proportional to real part of the Fourier Transform and have been modelled by the following expression:

$$S(\nu) = A.\Re\, FT\left\{J_2\left(mt.\,sinc\left(\frac{\omega_m t}{2}\right)\right).\cos(\omega_m t).\,\phi(t)\right\}$$

(3)

Where $J_2(x)$ is the second order Bessel function, $m$ the FM depth, $\omega_m$ the modulation rate, and $\phi(t)$ is the correlation function and corresponds to the desired line profile model. The time functions were constructed with a time step equal to the inverse of the product of the number of data points and frequency step of the measured spectra. The real component of the FFT was multiplied by a scaling factor $A$ to account for the amplitude of the measured molecular signal centered around zero. This factor is dependent on numerous aspects. Not only does it account for the responsivity of the instrument including elements like the detector sensitivity and amplifier gains, but it will depend on the line strength and the mixing ratio of trace gas. For a given molecular line all of these contributions to $A$ remain constant with the exception of the mixing ratio. The correlation function used is given below, it corresponds to a Voigt profile representing the convolution of a Gaussian profile originating from the Doppler effect, with a Lorentzian from the collisional relaxation.

$$\phi(t) = exp\left[i\nu_0 t - 2\pi\Delta\nu_L t - \left(\frac{\pi\Delta\nu_D t}{ln(2)}\right)^2\right]$$

(4)

The center frequency is denoted by $\upsilon_0$, while $\Delta\upsilon_L$ and $\Delta\upsilon_D$ are the collisional and Doppler linewidths respectively. The molecular signal was observed in the presence of a baseline signal from the Fabry-Perot effect of the measurement cell. It was easily removed by applying polynomial function (degree 2 to 4), and was established based on two zones of the spectra located above and below the targeted line that was centrally positioned. Following this initial step, a standard minimization function (Matlab environment) was used to fit the experimental data with the model described by equations 3 and 4. During the fitting the amplitude $A$, center frequency $\upsilon_0$, and collisional linewidth $\Delta\upsilon_L$ are free. All other parameters where set to their values corresponding to the experimental conditions. The concentration of $H_2S$ of a given sample was determined by the ratio of the amplitude parameter $A$, of the sample to that of the calibration gas. The fitting procedure determined the confidence interval of the fitted parameters to one standard deviation, these values were used to estimate the uncertainty of the measured concentration. The uncertainty resulting from the pressure gauge and temperature stability is much less compared that estimated by the fitting confidence interval.

## 6. Results and Discussion

To correctly validate the potential of the THz instrument to probe the freshness of salmon, samples with storage times varying from 5 to 13 days were examined and the comparison between THz spectroscopy and SIFT-MS was performed on the same sample simultaneously. In a previous study [30], exponential increase of $H_2S$ concentration and decrease in the sensory quality were detected from days 5 to 6 for comparable storage conditions; experienced panelists generally recognized samples with $H_2S$ concentrations exceeding 400-500 ppb to have low sensory quality (score 1-2 on a five-point scale, where 1=spoiled and 5=fresh). Hence, the detection of less than 500 ppb $H_2S$ by THz spectroscopy was considered important in the present study. However, it must be emphasized that the sensory quality is affected by the whole perceived profile of headspace trace gases and that different samples with the same storage times will not necessarily produce the same trace concentration [30]. Comprehensive studies of quality degradation require a large degree of sample duplication in order to ensure a significant statistical representation; this was not the objective of this work which aimed at demonstrating the potential of THz spectroscopy to probe the freshness of a given sample.

Example spectra for the strongest and weakest concentrations are shown in figure 4. The molecular signal is observed in the presence of a strong baseline variation due to the Fabry-Perot effect of the measurement cell. The polynomial baseline and fitted model display an excellent correspondence with the measured data. At day 13, the concentration measured by THz spectroscopy $C_{THz}$ = 3870 ± 100 ppb is in full agreement with the value measured by SIFT-MS for the same sample $C_{SIFT\text{-}MS}$ = 3880 ± 130 ppb. The available SNR on the molecular signal here is evaluated to be 24, defined as the ratio of fitted amplitude to the standard deviation of the residue. As expected, a much weaker trace was measured for a storage time of 5 days; the THz measurement $C_{THz}$ = 440 ± 80 ppb is compared to $C_{SIFT\text{-}MS}$ = 470 ± 80 ppb. However, the discrepancy of 30 ppb is well within the measurement uncertainties of both techniques. The SNR of the THz measurement is 2, indicating that the limit of detection (LOD) for this configuration is considered to be in the region of 220 ppb.

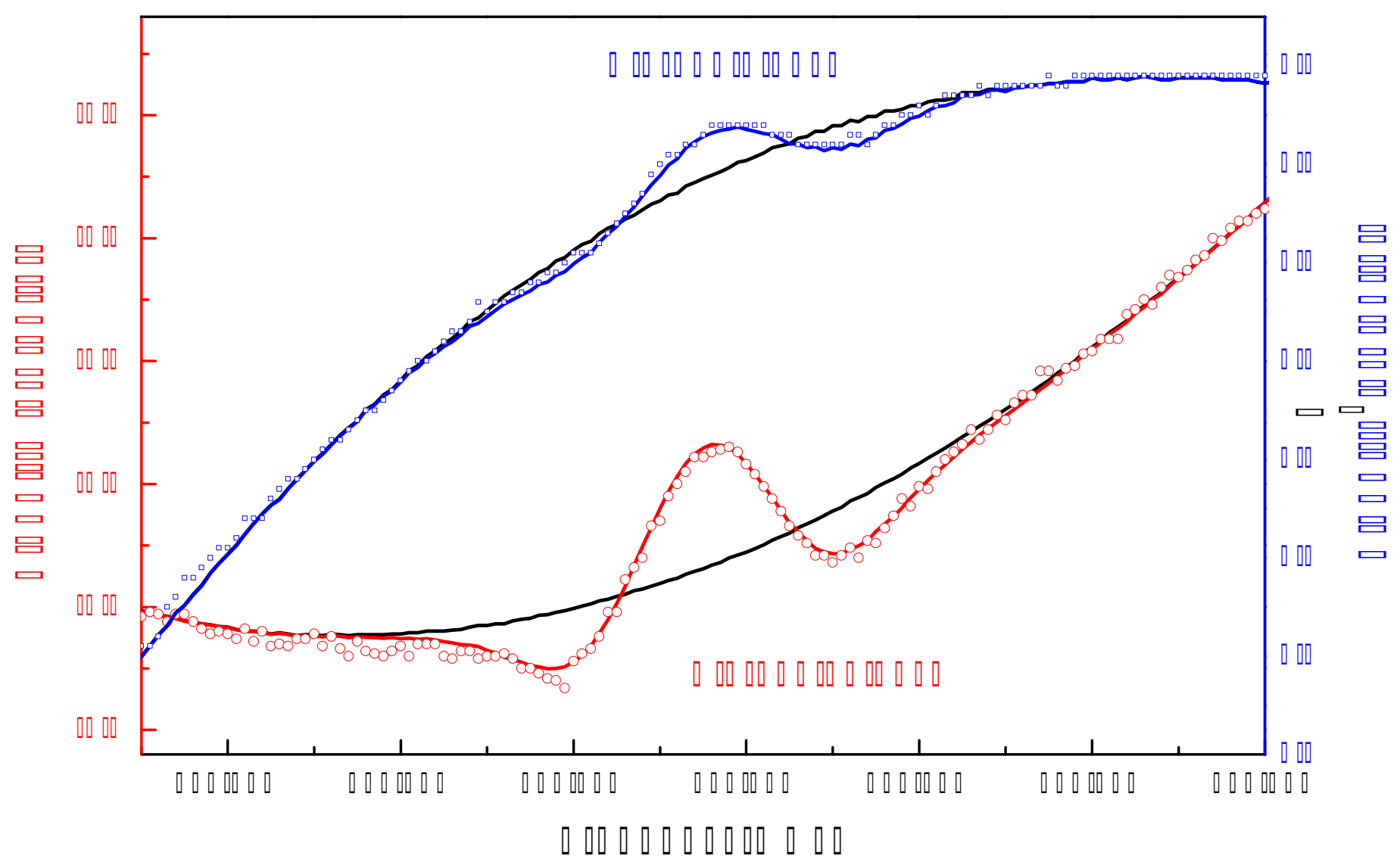

**Figure 4. Measured spectrum of $H_2S$ in the headspace gas of Atlantic salmon fillet portions stored for 13 days (red) and 5 days (blue) under 100% $N_2$ at 4 °C, modulation depth of 1188 kHz, interaction length of 1.128 m, temperature of 293 K, pressure 0.25 mbar, total measurement time ca. 10 minutes, integration time of a single point 1 s. Recorded spectra (circles or squares), fitted model (colored lines), polynomial baseline (black line). The vertical scale used for the 5 day sample has been increased by a factor x4 to aid viewing (new scale in blue on the right, original scale is shown in red on the left).**

A direct comparison of the quantification of the individual samples by THz spectroscopy and SIFT-MS was made by plotting $C_{THz}$ as a function of $C_{SIFT\text{-}MS}$, figure 5. The parameters show a strong correlation with a coefficient of R = 0.996, while a weighted linear fit of $C_{THz}$ yielded an offset of 15 ± 33 ppb and a scaling factor of 1.00 ± 0.03. The offset is well within the measurement uncertainties and the scaling factor of 1.00 indicates that there were no systematic measurement differences. The quantification of ppb-range hydrogen sulfide concentrations by THz spectroscopy was thus successfully validated by the SIFT-MS technique.

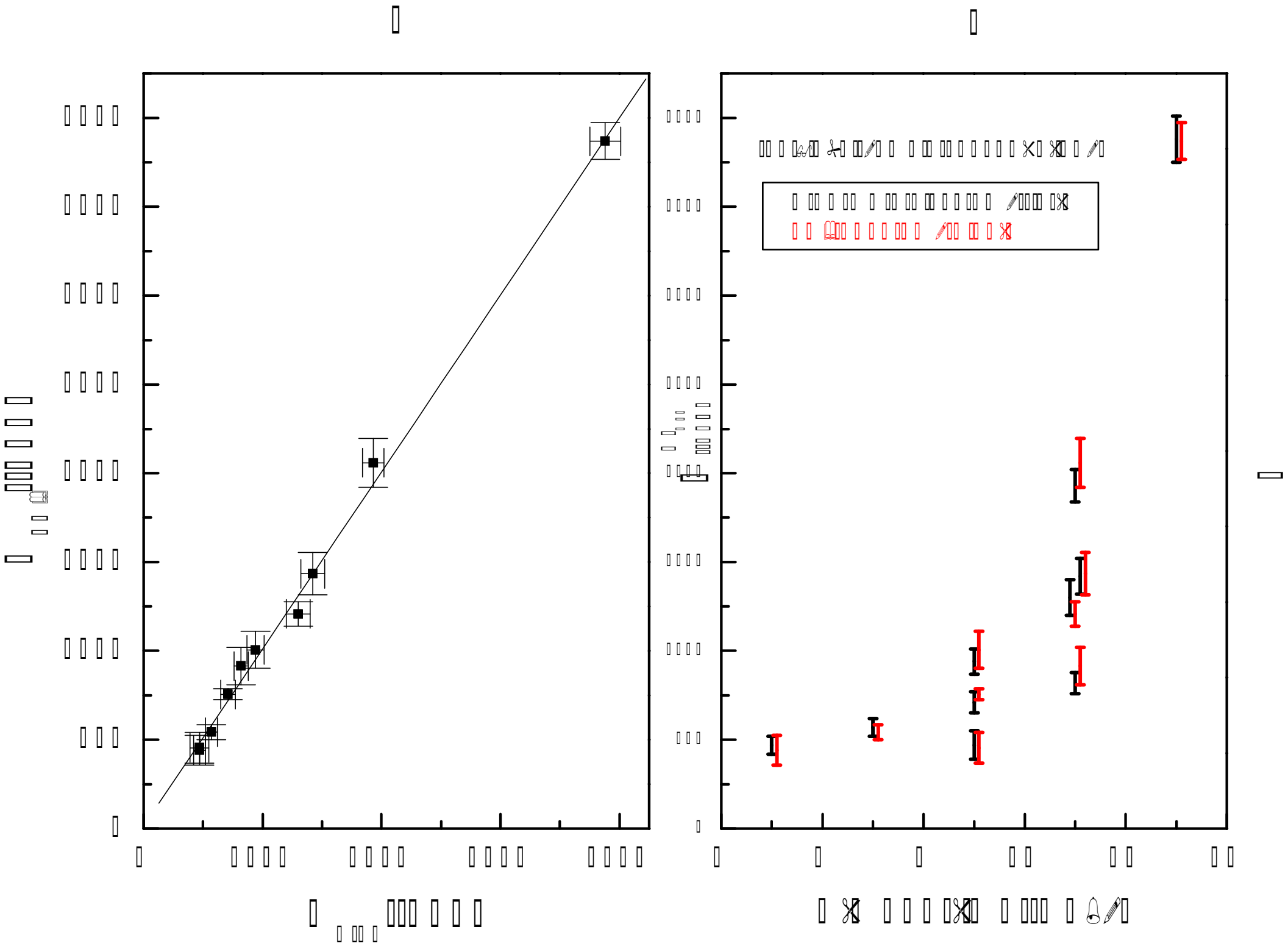

**Figure 5. Left pane: concentration of $H_2S$ measured by THz spectroscopy as a function of concentration measured by SIFT-MS for each sample (weighted linear fit is shown by the black line). Right pane: measured concentration values as a function of storage time for each sample. Red bars THz, black bars SIFT-MS. Length of bars (red THz, black SIFT-MS) indicate the measurement uncertainty. To aid the visualization, some of the storage times have been artificially adjusted by adding 0.1 to avoid overlapping.**

The examination of the values for each sample as a function of storage time is given in the right pane of figure 5. This more clearly indicates the dispersion of the concentration values for samples with the same storage times. Indeed, inspection of the data confirms that the clusters on the figures are the THz and SIFT-MS measurements of the same sample. For example, for a storage time of 9 days, three different concentrations values are clearly obtained. This is also the case for the four samples with a storage time of 11 days, which was due to the rapid development of $H_2S$ by the samples in this time window. The values for the individual samples were corroborated by both techniques, hence the red and black bars representing the same sample are close together. Furthermore, the measurement uncertainties were smaller than the observed sample dispersion.

The SIFT-MS instrument is able to monitor a large variety of trace gases as shown in Table A. With the exception of carbon disulfide $CS_2$, all these compounds are polar and therefore potentially detectable using gas phase rotational spectroscopy. Some of these molecules such as hydrogen sulfide, acetone, ethanol and ammonia have well understood THz signatures that have been comprehensively studied due to their detection in the interstellar medium. Accurate listings of the THz line frequencies and intensities of their rotation spectra may be calculated and are available in molecular spectroscopic databases such as JPL or CDMS [26], [32]. The compounds such as dimethyl amine, trimethyl amine, dimethyl sulfide, methyl mercaptan ethyl acetate, acetic acid and butanone form a second category of trace gases : these molecules are not present in the spectroscopic databases but their rotational spectroscopy has already been studied at microwave/millimeter wavelengths and the available molecular parameters may be used to simulate the rotational fingerprints in these spectral domains. The aforementioned compounds are listed in Table B with the reference to their rotational studies. Our ability/inability to detect these potential spoilage indicators depends on the absorption of the a->b rotational transition which in a first approximation may be quantified by the following equation (5) for the absorption coefficient [26]:

$$\alpha(\nu) = 1.3384 * 10^{-23} \nu_{ab} S_{ab} e^{-\frac{E_a}{kT}} (1 - e^{-\frac{\nu_{ab}}{kT}}) \frac{N\mu^2}{Q(T)} \tag{5}$$

with N: the number of absorbing molecules in molec.cm$^{-3}$, $\nu_{ab}$: the transition frequency in MHz, $S_{ab}$: the line strength, $\mu$: the permanent dipole moment component involved in the transition in Debye and *Q(T)* the partition function. The last part of the equation shows that the absorption is not only proportional to the concentration N, but also to the square of the permanent dipole moment $\mu^2$, and the inverse of the partition function $\frac{1}{Q(T)}$. The permanent dipole moment [33] and the rotational partition function [34] at T=300K calculated from the rotational constants are included in Table B for each molecule. While the dipole moments are rather similar, ranging from 0.6 Debye for trimethyl amine to 2.9 Debye for acetone, the room temperature rotational partition functions vary by several orders of magnitude. The lighter molecules ammonia ($NH_3$) and $H_2S$ have values close to 200, whereas heavier species such as ethyl acetate or butanone have values in excess of two orders of magnitude larger. For our application, the number of molecules available for sensing is directly linked to the concentration measured as a function of time by SIFT-MS (Table B).

| Molecule | | SIFT-MS concentrations | | | | | | Dipole moment | Rotational partition function | Detectability ratio | Ref |
|---|---|---|---|---|---|---|---|---|---|---|---|
| | | Day 5 *ppb* | Day 7 *ppb* | Day 9 *ppb* | Day 11 *ppb* | Day 13 *ppb* | Uncertainty *ppb* | *Debye* | | | |
| hydrogen sulfide | $H_2S$ | 470 | 650 | 650 | 1370 | 3880 | 80 | 0.974 | 254 | 17.619 | [26] |
| acetone | $CH_3COCH_3$ | 70 | 91 | 83 | 105 | 112 | 25 | 2.930 | 42492 | 0.028 | [26] |
| ethanol | $C_2H_5OH$ | 2200 | 7700 | 11000 | 17600 | 30000 | 500 | 1.462 | 17003 | 4.589 | [26] |
| ammonia | $NH_3$ | 40 | 41 | 39 | 40 | 37 | 9 | 1.472 | 215 | 0.248 | [26] |
| dimethylamine | $CH_3NHCH_3$ | 34 | 47 | 47 | 52 | 110 | 16 | 1.010 | 17094 | 0.008 | [35] |
| dimethylsulfide | $CH_3SHCH_3$ | 90 | 170 | 250 | 290 | 1050 | 40 | 1.554 | 31448 | 0.098 | [36] |
| ethyl acetate | $C_4H_8O_2$ | 60 | 120 | 170 | 220 | 340 | 40 | 2.130 | 158465 | 0.012 | [37] |
| methyl mercaptan | $CH_3SH$ | 30 | 120 | 250 | 850 | 2960 | 50 | 1.520 | 6826 | 1.219 | [38] |
| acetic acid | $CH_3COOH$ | 33 | 52 | 53 | 54 | 33 | 23 | 1.700 | 36621 | 0.003 | [39] |
| butanone | $C_4H_8O$ | 9 | 14 | 19 | 48 | 1069 | 18 | 2.779 | 90028 | 0.112 | [40] |
| trimethylamine | $C_3H_9N$ | 24 | 30 | 25 | 30 | 46 | 11 | 0.612 | 44921 | 4.637E-04 | [41] |

**Table B. Concentrations of various trace gases (mean values for multiple samples with identical storage time) as a function of storage time, quantified by SIFT-MS in the headspace of Atlantic salmon stored under 100 % $N_2$ at 4 °C. The dipole moments [33] are provided with the calculated rotational partition functions. They are used to estimate suitability as a spoilage indicator by means of the detectivity ratio defined in equation (6).**

Overall, the different concentrations vary over more than four orders of magnitude if we compare the concentration of butanone measured on the 5$^{th}$ day and the concentration of ethanol on the 13$^{th}$ day (Table B). In order to identify a wider range of spoilage indicators, a comparison relative to $H_2S$ has been made. The three essential parameters concentration, dipole moment and partition function have been taken into account for each molecule. A detectability ratio $R_j$ has been calculated for each molecular component $j$, and the experimental detection limit of $H_2S$ ($N_{LOD}$ = 220 ppb) is compared with the typical value for the given species on day 13 using the following relation:

$$R_j = \frac{\left[N_{13days}\, \mu^2 / Q_{300K}\right]_j}{\left[N_{LOD}\, \mu^2 / Q_{300K}\right]_{H_2S}} \tag{6}$$

Firstly, we can notice that the detectability is favourable to $H_2S$ compared to the other molecules, which also explains why this molecule was selected for this demonstration. Secondly, comparing the $R_j$ values of the different freshness tracers, three categories of molecules may be identified considering a potential detection THz:

- Ethanol, methyl mercaptan and ammonia should be detectable with a sensitivity level similar to $H_2S$ using the existing system.
- The detection of dimethyl sulfide, acetone and butanone may be expected with a sensitivity improvement of one or two orders of magnitude.
- Spoilage indicators such as dimethyl amine, trimethyl amine, ethyl acetate or acetic acid are not detectable with the sensitivity of today's THz instrumentation.

## 7. Conclusion

Despite monitoring of food spoilage being a key challenge for today's society, at present there are no widely employed solutions able to assess the freshness of packed food without opening the package. Since THz waves will interact with many trace gases produced by microbial activity occurring during food storage, they may be employed to perform the analysis of the headspace gas of packed foodstuffs. In the present study, we have demonstrated the potential of THz waves to monitor the production of hydrogen sulfide during refrigerated storage of Atlantic salmon under 100 % $N_2$. The existing instrument may be applied to a wider range of molecular components which may be of interest in order to quantify multiple spoilage indicators of salmon or to widen the scope to other food products. Standard techniques routinely employed in the infrared such as multiple pass cells or cavity enhanced absorption spectroscopy should provide an increased sensitivity to this end. Along with other published investigations [21], [22], this present comprehensive comparison paves the way towards a new solution for the analysis of gases. Although it promises the required selectivity and quantification, its shortcoming is the present

Technology Readiness Level (TRL) restricting access to specialized laboratories. The integration of THz waveguides [42] and resonators [43] using silicon-based fabrication processes presents a longer-term perspective to construct a sensor operating on this principle that can be placed inside the package.

**Conflicts of interest**

There are no conflicts to declare

**Acknowledgement**

This study was undertaken as part of the Terafood project which is financially supported by the European Regional Development Fund and the province Oost-Vlaanderen.